\documentclass[12pt,a4paper]{article}
\usepackage[T1]{fontenc}
\usepackage[utf8]{inputenc}
\usepackage{lmodern}
\usepackage{amsmath,graphicx,geometry,csquotes}
\usepackage{hyperref} 
\usepackage{url}

\title{Mastering Uncertainty:\\From Understanding to Prediction \footnote{chapter to be published in a special volume as a contribution to
the Spring 2025 lecture series ``Uncertainty: Navigating the Unpredictable in Society, Cognition, and Existence'' at the University of Zurich.}}
\author{Didier Sornette\\
Chair Professor, Risks-X at SUSTech, Shenzhen\\
Professor Emeritus, ETH Zurich}
\date{11 October 2025}

\begin{document}
\maketitle

\begin{abstract}
Uncertainty defines our age: it shapes climate, finance, technology, and society, yet remains profoundly misunderstood. We oscillate between the illusion of control and the paralysis of fatalism. This paper reframes uncertainty not as randomness but as ignorance: a product of poor models, institutional blindness, and cognitive bias. Drawing on insights from physics, complex systems, and decades of empirical research, I show that much of what appears unpredictable reveals structure near transitions, where feedbacks, critical thresholds, and early-warning signals emerge. Across domains from financial crises to industrial disasters, uncertainty is amplified less by nature than by human behavior and organizational failure. To master it, prediction must shift from prophecy to diagnosis, identifying precursors of instability rather than forecasting exact outcomes. I propose a framework of dynamic foresight grounded in adaptive leadership, transparent communication, and systemic learning. Mastering uncertainty thus means transforming fear into foresight and building institutions that navigate, rather than deny, the complexity of change.
\end{abstract}

\maketitle

\section{Introduction: The Paradox of Uncertainty}

Uncertainty is the inescapable companion of our age. We encounter it in the climate system, in financial markets, in pandemics, and in the very fabric of social and technological change~\cite{Wilson2002ScientificUncertainty,Khan2018EmbracingUncertainty,Crutchfield2020HiddenFragility}. Yet, despite its ubiquity, we often misunderstand its nature. We alternate between denial---pretending the world is predictable if only we collect enough data---and fatalism---believing that uncertainty makes foresight impossible. Both extremes are errors of framing~\cite{Faccia2025IllusionControl}.

My central claim is that much of what we call ``unpredictable'' is not inherently so. Uncertainty in many systems stems less from the intrinsic complexity of nature than from human-imposed constraints: our institutional blindness, cognitive biases, and misplaced questions~\cite{Emmons2018MitigatingCognitiveBiases,Anagnou2025InstitutionBootstrapping,Bland2012DifferentUncertainty}. When we learn to ask questions at the right level of description and recognize the signals of approaching change, uncertainty becomes not an enemy but a guide~\cite{Hullermeier2021AleatoricEpistemic}.

This essay translates a lifetime of work---across physics, finance, and risk management---into a broader reflection on uncertainty as both a challenge and an opportunity for society. Its structure addresses: (1) the conceptual vantage point---what kind of uncertainty we face; (2) the field of manifestation---how these uncertainties shape our world; and (3) the horizon of response---what we can do to live, act, and govern more intelligently in their presence.

\section{Conceptual Vantage Point: What Kind of Uncertainty?}

\subsection{From randomness to ignorance}

In everyday language, ``uncertainty'' is often equated with randomness, but the two are distinct. Randomness refers to intrinsic variability---the roll of a fair die or the decay of a radioactive atom. Uncertainty, instead, often reflects ignorance: limited knowledge, incomplete models, or poor framing of the question. We cannot specify all possible events, quantify their probabilities, or even define what ``loss'' means \cite{Hullermeier2021AleatoricEpistemic}. 

Our attempts to manage uncertainty often fail for five recurrent reasons:
\begin{enumerate}
    \item Lack of imagination to explore relevant scenarios.
    \item Lack of courage or incentives to confront real problems.
    \item Misguided reward systems that prize short-term success over learning.
    \item Lack of understanding of complex systems and their feedbacks.
    \item Failure of leadership and communication to share critical information.
\end{enumerate}

These are not technical failures but conceptual ones. They define what I call the \emph{lamp-post problem}: like the drunk searching for his keys under the streetlight because that is where the light is, we focus our analyses where the data are easy to collect rather than where the truth is likely to lie \cite{RoyZeckhauserGrapplingIgnorance}.

\subsection{Illusions of complexity}

Complexity is frequently invoked to justify our ignorance. In the early 2000s, Stephen Wolfram popularized the idea of ``computational irreducibility'': some systems, such as certain cellular automata (e.g., Rule 110), are so intricate that their outcomes can be known only by simulating every step \cite{Wolfram2002_NKS}. This idea became a metaphor for unpredictability in nature.

Yet this message is misleading. Complexity and unpredictability depend on the level of description. The chaotic motion of air molecules in this room is maximally complex, but at the right scale we obtain the beautifully simple ideal gas law, \(pV = nRT\). The art lies in ``coarse-graining'', i.e,, choosing the proper level at which patterns emerge. What seems unpredictable at one scale becomes regular and even predictive at another \cite{IsraeliGoldenfeld2006}.

This ``renormalization'' logic, borrowed from physics, teaches that irreducibility is not absolute. It reflects our failure to identify the relevant variables and control parameters. Many uncertainties vanish once we understand the structure of interactions and the proximity to transitions.

\subsection{Approaching transitions: the physics of change}

In systems as diverse as ecosystems, economies, or social networks, large transitions occur when a small change in one control parameter---temperature, leverage, trust---pushes the system beyond a critical threshold. Near such \emph{bifurcations}, sensitivity to perturbations increases dramatically. Recovery from shocks slows down, correlations rise, and fluctuations amplify \cite{Scheffer2012_EWS}.

These are universal \emph{early-warning signals}. They do not predict the precise moment of transition, but they reveal that a regime shift is approaching. The same mathematics applies to ecosystems approaching collapse, brains nearing epileptic seizures, and markets inflating speculative bubbles \cite{Sornette2003_CriticalCrashes,{Sornette2006_CriticalPhenomena}}. The universality of these signatures shows that uncertainty has structure.

\subsection{The mirage of unknowability}

Many influential figures in economics have declared crises ``unpredictable.'' From Keynes's 1921 assertion that the future is inherently uncertain \cite{Keynes1921_Treatise} to Bernanke's resignation that ``we have no way of knowing when the next crisis will occur,'' \cite{Bernanke2010_ImplicationsCrisis} this narrative has justified reactive policymaking. Yet empirical research contradicts it. The great financial crashes---1929, 1987, 2000, 2008---show distinct precursory patterns: accelerating growth, often decreasing or stable volatility, and growing synchronization across assets \cite{JSL99,JohansenLedoitSornette1998_CrashesCriticalPoints}. 

The failure is not one of prediction, but of \emph{recognition}. Policymakers and institutions, trapped in models assuming equilibrium and linear response, overlook the telltale signs of positive feedback and collective amplification \cite{IllusionSC14}.

\section{Field of Manifestation: Where Uncertainty Matters}

\subsection{Finance: the myth of black swans}

Financial markets are often portrayed as ruled by ``black swans'', defined as unforeseeable outliers. The analyses performed by my collaborators and I over many years suggest otherwise. Large drawdowns, when measured properly as sequences of cumulative losses rather than isolated daily returns, form a distinct statistical class. I call these events ``dragon-kings'': rare, extreme, but generated by identifiable mechanisms of self-amplification \cite{Sornette2009_DragonKings, SornetteOuillon2012_DragonKings}.

In this view, bubbles and crashes are not random accidents but phases of endogenous instability driven by herding, leverage, and reflexivity. The same positive feedbacks that fuel exuberant growth make the system fragile. Predicting the \emph{timing} of such instabilities requires monitoring the degree of endogeneity, growth acceleration, and collective coupling \cite{Sornette2009_DragonKings}.

This framework has achieved real-world validation. My research group has successfully issued advance warnings before the 2006 U.S. real-estate peak \cite{zhousorpredireal} and major Chinese stock market corrections \cite{Chinapred}. The lesson is simple: crises are the price of ignoring the dynamics of our own collective behavior.

\subsection{Engineering and industry: when risk information fails}

The same logic applies beyond finance. Together with Dmitry Chernov, we examined over 500 cases of industrial, environmental, and technological disasters, from Bhopal to Chernobyl, from the Challenger explosion to Fukushima. Despite their diversity, these events share strikingly similar roots: \emph{risk information concealment}. Signals of danger existed but were suppressed by organizational culture, misaligned incentives, and fear of blame \cite{ChernovSornette2016_Catastrophes}.

We identified more than thirty recurring causes, ranging from ``no bad news'' cultures and short-term profit pressures to weak communication channels and regulatory capture. The 1986 Challenger disaster remains a textbook example: engineers had data showing O-ring damage increased sharply with lower temperatures, but management ignored the trend. The shuttle launched at $2\,^{\circ}\mathrm{C}$ and exploded.

These failures illustrate that uncertainty is not only cognitive but institutional. Systems become blind not because information is absent but because it cannot circulate. This blindness is self-generated and therefore, in principle, correctable.

\subsection{Societal scale: crises as social amplifiers}

At the societal level, uncertainty interacts with psychology and media dynamics. In modern democracies, fear is often instrumentalized---whether the fear of financial collapse, pandemics, or geopolitical threats. This \emph{manufactured uncertainty} narrows public imagination and impairs resilience. Paradoxically, while institutions claim to manage risk, they often amplify collective anxiety, creating cycles of overreaction and distrust \cite{Kasperson1988_SocialAmplificationRisk}.

Recognizing this mechanism is essential to restoring agency. Societies that conflate uncertainty with danger become risk-averse and stagnant; those that learn to interpret uncertainty as information become innovative and adaptive.

\section{Horizon of Response: How to Live and Govern with Uncertainty}

\subsection{From passive to dynamic management}

The traditional approach to risk management is static: assess probabilities, compute expected losses, and design controls. But this assumes stationarity---the idea that the underlying processes do not change. In reality, most systems of interest are \emph{non-stationary}: their very rules evolve with time. Managing them is less like building a fortress than like piloting a motorcycle on a mountain road \cite{Khan2016_DynamicRiskAssessment}.

I often use the motorcyclist metaphor to capture the essence of \emph{dynamic risk management}: continuously sensing, anticipating, and adjusting. The goal is not to eliminate uncertainty but to navigate it by maintaining stability through adaptation. This stands in sharp contrast to \emph{static risk management}, which consists in counting on the helmet, protective clothing, and gloves---the passive defences designed for the moment of impact. But if I ever find myself relying on them, it means the essential has already failed and I am dead. In real life as in complex systems, survival depends not on protection during and after the crash, but on constant awareness and correction before it.

\subsection{Prediction as diagnosis, not prophecy}

Prediction in complex systems should not mean forecasting exact outcomes, but identifying the conditions under which change becomes likely. In medicine, a competent doctor practicing personalised medicine does not predict the precise moment a patient will fall ill but recognizes precursors---symptoms, risk factors, and feedbacks---that demand intervention. Similarly, in finance or environmental policy, prediction means \emph{diagnosing instability} \cite{Soleimani2021_DiagnosticsComplexSystems}.

This shift---from prediction as prophecy to prediction as diagnosis---reconciles the human need for foresight with the reality of complexity. It restores meaning to the concept of scientific prediction without the hubris of omniscience.

\subsection{Constructive uncertainty and innovation}

Uncertainty also has a creative dimension. Innovation, entrepreneurship, and evolution itself rely on exploring the unknown. Regime shifts are not merely crises but opportunities for renewal. The key is to distinguish \emph{constructive} uncertainty, which opens new pathways, from \emph{destructive} uncertainty, which results from opacity and miscommunication.

For this, organizations must cultivate an ecology of risk information---where anomalies are reported, dissent is valued, and leadership rewards transparency. The empirical work of my collaborators and myself shows that companies and institutions that tolerate ambiguity, rather than suppressing it, are more resilient and innovative \cite{ChernovSornette23_Catastrophes}. 

\subsection{Recommendations for resilient systems}

Across domains---from finance to engineering, from organizational failures to social crises---the decisive factor separating collapse from resilience is \emph{leadership}.  
Resilience is not a property of materials or algorithms but of human collectives guided by leaders who can interpret weak signals, foster trust, and act before systems drift beyond control \cite{Sornette2006_CriticalPhenomena,Schuttner2021_OrgResilience}.  
From my decades of research, a set of interlocking principles emerge, synthesizing the lessons of prediction, communication, and adaptation developed throughout this essay.

\begin{enumerate}
    \item \textbf{Lead from foresight, not hindsight.} The essence of governance is anticipation. True leaders act as dynamic navigators, constantly scanning for early-warning signals and adapting course before crises unfold---as a motorcyclist reads the road rather than relying on the helmet. Waiting for certainty is the surest path to disaster.
    
    \item \textbf{Build a culture of fearless communication.} Uncertainty becomes lethal when information cannot flow. Leaders must make it safe---and rewarding---for people to voice uncomfortable truths. Silence and concealment are the invisible accelerants of catastrophe \cite{Edmondson2019_FearlessOrganization,ChernovSornette23_Catastrophes}.
    
    \item \textbf{Replace blame with systems learning.} Every accident, from the Challenger to Fukushima, reveals how institutions personalize error instead of tracing the systemic pathways that produced it. The mature organization investigates causes, not culprits \cite{Reason1997_ManagingRisk, Dekker2018_JustCulture}.
    
    \item \textbf{Align vision with action.} Credibility is the currency of resilience. When leaders speak of safety, sustainability, or foresight, their decisions and incentives must embody those words. Otherwise, communication decays into noise, and trust---the most fragile form of capital---is lost \cite{CCL2024_WhyTrustMatters}.
    
    \item \textbf{Value stability over speed.} Markets, institutions, and technologies are rewarded for acceleration; yet, every sustainable system depends on mechanisms of negative feedback that temper growth. Wise leaders know when to slow down. Stability is not stagnation; it is the condition for creative longevity.
    
    \item \textbf{Institutionalize reflection and imagination.} The greatest failures stem from a failure of imagination---from not asking the right questions. Leadership must therefore cultivate collective imagination: scenario planning, dissent, and long-term thinking that extends beyond immediate metrics. 
\end{enumerate}

Resilient leadership is thus a synthesis of \emph{vigilance, humility, and imagination}.  
It transforms uncertainty from a threat into a source of continuous learning.  
To master uncertainty is not to suppress it, but to organize around it---to create institutions that breathe with change rather than shatter under it.

\subsection{The future of prediction: maps, not forecasts}

In the 19th century, \'Emile de Girardin wrote, ``To govern is to foresee.'' But foresight today must be understood dynamically. Rather than static predictions, we need {\rm maps}---representations of the landscape of possible futures and the bifurcations that separate them. By identifying zones of fragility and resilience, such maps guide navigation through uncertainty.

My ongoing work at the Financial Crisis Observatory \cite{FCO1}-\cite{FCO4} and the Risks-X Institute (\url{https://risks-x.sustech.edu.cn}) with students, post-docs and collaborators develops such dynamic mapping tools, combining big data with theories of phase transitions. The same principles apply to planetary health, energy transitions, and social innovation: the challenge is not to foresee every event but to understand the logic of emergence.

\section{Conclusion: From Fear to Foresight}

The 21st century confronts us with intertwined crises---climate instability, financial fragility, technological disruption---that feed on our collective blindness. Yet these crises also reveal a deeper truth: uncertainty is not the opposite of knowledge but its frontier. Mastering uncertainty means transforming ignorance into structure, and fear into preparation.

We must abandon both the illusion of full control and the resignation of fatalism. Between these lies a new epistemology: one that views uncertainty as an invitation to refine our models, rethink our institutions, and renew our imagination. 

Every catastrophe teaches the same lesson: nature, society, and technology are more imaginative than we are. The task of science and policy alike is to catch up---to illuminate not only where the light already shines, but also where the keys to our future truly lie.


\begin{thebibliography}{99}

\bibitem{Wilson2002ScientificUncertainty}
J.A. Wilson, Scientific Uncertainty, Complex Systems, and the Design of Common-Pool Institutions, in The Drama of the Commons (National Research Council, 2002), Eds., Paul Stern, Elinor Ostrom, Thomas Dietz, and Nives Dolsak.  

\bibitem{Khan2018EmbracingUncertainty}
S.~Khan, J.~Vaillancourt Rosenau, and J.~Reiss, Embracing uncertainty, managing complexity: applying complexity thinking principles to transformation efforts in healthcare systems, BMC Health Services Research, 18, 192 (2018).  

\bibitem{Crutchfield2020HiddenFragility}
J. P. Crutchfield, The Hidden Fragility of Complex Systems--Consequences of Change, Changing Consequences, 
Cultures of Change: Social Atoms and Electronic Lives, G. Ascione, C. Massip, J. Perello, editors, 
ACTAR D Publishers, Barcelona, Spain, 98-111 (2009).
(arXiv preprint arXiv:2003.11153)  

\bibitem{Faccia2025IllusionControl}
A. Faccia, P. Petratos, and F. Manni, The Illusion of Control: How Knowledge and Expertise Misclassify Uncertainty as Risk, Risks 13 (10): 188 (2025).  

\bibitem{Emmons2018MitigatingCognitiveBiases}
D.L . Emmons, T.A. Mazzuchi, S. Sarkani and C.E. Larsen,  Mitigating cognitive biases in risk identification:  Practitioner Checklist for the aerospace sector, Def. Acquis. Res. J. (25 (1), 52-93 (2018).

\bibitem{Bland2012DifferentUncertainty}
A.R. Bland and A. Schaefer, Different Varieties of Uncertainty in Human Decision-Making,
Front. Neurosci. 6, 85, pp. 1-11 (2012).

\bibitem{Anagnou2025InstitutionBootstrapping}
S. Anagnou, C. Salge, and P. R. Lewis, Uncertainty, bias and the institution bootstrapping problem, arXiv preprint arXiv:2504.21579 (2025).  

\bibitem{Hullermeier2021AleatoricEpistemic}
E. H\"ullermeier and W. Waegeman, Aleatoric and Epistemic Uncertainty in Machine Learning: An Introduction to Concepts and Methods,  
Mach. Learn. 110, 457-506 (2021).

\bibitem{RoyZeckhauserGrapplingIgnorance}
D. Roy and R. Zeckhauser, Grappling with Ignorance: Frameworks from Decision Theory, J. Benefit Cost Anal. 6 (1), 33-65 (2015).  

\bibitem{Wolfram2002_NKS}
S. Wolfram, A New Kind of Science, Wolfram Media, 2002.  

\bibitem{IsraeliGoldenfeld2006}
N. Israeli and N. Goldenfeld, Coarse-graining of cellular automata, emergence, and the predictability of complex systems, 
Physical Review E 73, 026203 (2006).  

\bibitem{Scheffer2012_EWS}
M. Scheffer, J. Bascompte, W. A. Brock, V. Brovkin, S. R. Carpenter, V. Dakos, H. Held, E. H. van Nes, M. Rietkerk, and G. Sugihara, Early-warning signals for critical transitions, Nature 461, 53-59 (2009).  

\bibitem{Sornette2003_CriticalCrashes}
D. Sornette, Critical market crashes, Physics Reports 378 (1), 1-98 (2003).  

\bibitem{Sornette2006_CriticalPhenomena}
D. Sornette, Critical Phenomena in Natural Sciences: Chaos, Fractals, Self-organization and Disorder, Concepts and Tools, 2nd ed., Springer (2006).  

\bibitem{Keynes1921_Treatise}
J. M. Keynes, A Treatise on Probability, Macmillan, 1921.  

\bibitem{Bernanke2010_ImplicationsCrisis}
B. S. Bernanke, Implications of the Financial Crisis for Economics, Speech at the Conference Sponsored by the Center for Economic Policy Studies and the Bendheim Center for Finance, Princeton University (September 24, 2010).  

\bibitem{JSL99} 
A. Johansen, D. Sornette and O. Ledoit, Predicting Financial Crashes using discrete scale invariance,
Journal of Risk 1 (4), 5-32 (1999).

\bibitem{JohansenLedoitSornette1998_CrashesCriticalPoints}
A. Johansen, O. Ledoit and D. Sornette, Crashes as critical points, International Journal of Theoretical and Applied Finance 3 (2),  219-255 (2000).

\bibitem{IllusionSC14}
D. Sornette and P. Cauwels, 1980-2008: The Illusion of the Perpetual Money Machine and what it bodes for the future,
Risks 2, 103-131 (2014).

\bibitem{Sornette2009_DragonKings}
D. Sornette, Dragon-Kings, Black Swans and the Prediction of Crises, International Journal of Terraspace Science and Engineering 2(1), 1-18 (2009).  

\bibitem{SornetteOuillon2012_DragonKings}
D. Sornette and G. Ouillon, Dragon-kings: mechanisms, statistical methods and empirical evidence, European Physical Journal Special Topics 
205, 1-26 (2012).  

\bibitem{zhousorpredireal}
W.-X. Zhou and D. Sornette, Is There a Real-Estate Bubble in the US?
Physica A: Statistical Mechanics and its Applications 361, 297-308 (2006)
(prediction posted in June 2005 at http://arxiv.org/abs/physics/0506027)

\bibitem{Chinapred}
Didier Sornette, Guilherme Demos, Qun Zhang, Peter Cauwels, Vladimir Filimonov and Qunzhi Zhang,
Real-time prediction and post-mortem analysis of the Shanghai 2015 stock market bubble and crash,
Journal of Investment Strategies 4 (4), 77-95  (2015)

\bibitem{ChernovSornette2016_Catastrophes}
D. Chernov and D. Sornette, Man-made Catastrophes and Risk Information Concealment: Case Studies of Major Disasters and Human Fallibility, Springer (2016).  

\bibitem{Kasperson1988_SocialAmplificationRisk}
R. E. Kasperson, O. Renn, P. Slovic, H. S. Brown, J. Emel, R. Goble, J. X. Kasperson and S. W. Ratick, The Social Amplification of Risk: A Conceptual Framework, Risk Analysis, 8(2), 177-187 (1988).  

\bibitem{Khan2016_DynamicRiskAssessment}
F. Khan, S.J. Hashemi, N.Paltrinieri, P. Amyotte, V. Cozzani and G. Reniers, 
Dynamic risk management: a contemporary approach to process safety management, 
Current Opinion in Chemical Engineering 14,  9-17 2016).

\bibitem{Soleimani2021_DiagnosticsComplexSystems}
M. Soleimani, F. Khan, and P. Davari, Diagnostics and prognostics for complex systems: A review of methods and challenges, Quality and Reliability Engineering International 37 (8), 3746-3778 (2021).  

\bibitem{ChernovSornette23_Catastrophes}
Dmitry Chernov, Ali Ayoub, Giovanni Sansavini and Didier Sornette,
Averting disaster before it strikes (how to make sure your subordinates warn you
while there is still time to act), Springer (2023)

\bibitem{Schuttner2021_OrgResilience}
L. Sch\"uttner, K. Coleman, J. Ralston and M. Parchman,
The role of organizational learning and resilience for change in building quality improvement capacity in primary care, 
Health Care Manage Rev. 46 (2), E1-E7 (2021).

\bibitem{Edmondson2019_FearlessOrganization}
A. C. Edmondson, The Fearless Organization: Creating Psychological Safety in the Workplace for Learning, Innovation, and Growth, Hoboken, NJ: John Wiley \& Sons (2018).

\bibitem{Reason1997_ManagingRisk}
J. Reason, Managing the Risks of Organizational Accidents, London, Routledge(1997).  

\bibitem{Dekker2018_JustCulture}
S. Dekker, Just Culture: Restoring Trust and Accountability in Your Organization, 3rd Edition, CRC Press, London (2018).  

\bibitem{CCL2024_WhyTrustMatters}
Center for Creative Leadership, Why Leadership Trust Is Critical, Especially in Times of Change, CCL (2024).  

\bibitem{FCO1}
Didier Sornette, Ryan Woodard, Maxim Fedorovsky, Stefan Riemann, Hilary Woodard, Wei-Xing Zhou 
(The Financial Crisis Observatory)
The Financial Bubble Experiment: advanced diagnostics and forecasts of bubble terminations (2009)
(http://arxiv.org/abs/0911.0454)

\bibitem{FCO2} Didier Sornette, Ryan Woodard, Maxim Fedorovsky, Stefan Reimann,
 Hilary Woodard, Wei-Xing Zhou (The Financial Crisis Observatory),
 The Financial Bubble Experiment: Advanced Diagnostics and Forecasts of
 Bubble Terminations Volume II--Master Document (beginning of the experiment) (2010)
(http://arxiv.org/abs/1005.5675)

\bibitem{FCO3}
Didier Sornette, Ryan Woodard, Maxim Fedorovsky, Stefan Reimann,
 Hilary Woodard, Wei-Xing Zhou (The Financial Crisis Observatory),
 The Financial Bubble Experiment: Advanced Diagnostics and Forecasts of
 Bubble Terminations Volume II-Master Document (end of the experiment),
(http://arxiv.org/abs/1005.5675)

\bibitem{FCO4}
Ryan Woodard, Didier Sornette, Maxim Fedorovsky
The Financial Bubble Experiment: Advanced Diagnostics and Forecasts of
 Bubble Terminations, Volume III (beginning of experiment + post-mortem analysis) (2010)
 (http://arxiv.org/abs/1011.2882)
 

\end{thebibliography}
\end{document}